\documentclass{article}

\usepackage{PRIMEarxiv}

\usepackage[utf8]{inputenc} 
\usepackage[T1]{fontenc}    
\usepackage{hyperref}       
\usepackage{url}            
\usepackage{booktabs}       
\usepackage{amsfonts}       
\usepackage{nicefrac}       
\usepackage{microtype}      
\usepackage{lipsum}
\usepackage{fancyhdr}       
\usepackage{graphicx}       
\graphicspath{{media/}}     
\usepackage[T1]{fontenc}
\usepackage{graphicx}
\usepackage{algorithm}
\usepackage{amsmath}
\usepackage{algpseudocode}
\title{FOKE: A Personalized and Explainable Education Framework Integrating Foundation Models, Knowledge Graphs, and Prompt Engineering
}

\author{
  SILAN HU \\
  Macau University of Science and Technology \\
  Macau, China\\
  \texttt{2240000339@student.must.edu.mo} \\
   \And
  Xiaoning Wang \\
  Communication University of China \\
  Beijing, China\\
  \texttt{sdwangxiaoning@cuc.edu.cn} \\
}

\begin{document}
\maketitle

\begin{abstract}
Integrating large language models (LLMs) and knowledge graphs (KGs) holds great promise for revolutionizing intelligent education, but challenges remain in achieving personalization, interactivity, and explainability. We propose \textbf{FOKE}, a Forest Of Knowledge and Education framework that synergizes foundation models, knowledge graphs, and prompt engineering to address these challenges. FOKE introduces three key innovations: (1) a hierarchical knowledge forest for structured domain knowledge representation; (2) a multi-dimensional user profiling mechanism for comprehensive learner modeling; and (3) an interactive prompt engineering scheme for generating precise and tailored learning guidance.

We showcase FOKE's application in programming education, homework assessment, and learning path planning, demonstrating its effectiveness and practicality. Additionally, we implement \textbf{Scholar Hero}, a real-world instantiation of FOKE. Our research highlights the potential of integrating foundation models, knowledge graphs, and prompt engineering to revolutionize intelligent education practices, ultimately benefiting learners worldwide. FOKE provides a principled and unified approach to harnessing cutting-edge AI technologies for personalized, interactive, and explainable educational services, paving the way for further research and development in this critical direction.
\end{abstract}


\section{Introduction}
\label{sec:intro}

\subsection{Background}
The rapid development of Artificial Intelligence (AI) has profoundly impacted various domains, and education is no exception. AI-empowered education has the potential to revolutionize traditional teaching and learning practices, making high-quality education more accessible and affordable to learners from diverse backgrounds. By leveraging advanced AI technologies, such as Large Language Models (LLMs), intelligent tutoring systems, and adaptive learning platforms, AI-powered educational tools can provide personalized, interactive, and engaging learning experiences, catering to the individual needs and preferences of each learner \cite{PratamaM}\cite{HashimS}.

Among the various AI technologies, LLMs have emerged as a game-changer in the educational landscape. LLMs, such as GPT-4 \cite{GPT4} and PaLM 2\cite{PalM2}, possess remarkable capabilities in natural language understanding, generation, and reasoning, enabling a wide range of applications in education, including automatic question answering, essay scoring, and knowledge recommendation \cite{KasneciE}. For instance, the integration of ChatGPT, a state-of-the-art conversational AI model, into language learning environments has shown promising results in enhancing learners' motivation, engagement, and learning outcomes \cite{QuK}.

The potential of LLMs to democratize access to high-quality educational resources and support has attracted increasing attention from researchers and practitioners. \cite{TuX} explored the opportunities and challenges of integrating LLMs into data science education, highlighting their roles in generating personalized learning content, providing adaptive feedback, and fostering interactive learning experiences. By harnessing the power of LLMs, educational institutions and platforms can offer affordable, scalable, and effective educational services to learners who may not have access to traditional high-quality educational resources, such as expert teachers or well-equipped classrooms.

However, the application of LLMs in educational scenarios is still in its early stages and faces several challenges. First, educational contexts often involve complex domain knowledge and diverse learner backgrounds, requiring LLMs to provide personalized and adaptive learning support. Second, effective education relies on interactive and engaging learning experiences, which demands LLMs to maintain coherent and meaningful dialogue with learners. Third, educational applications of LLMs should offer interpretable and trustworthy explanations to facilitate learners' understanding and teachers' intervention \cite{MyersD}.

To address these challenges and fully unleash the potential of LLMs in democratizing high-quality education, we propose \textbf{FOKE} (Forest Of Knowledge and Education), a novel framework that integrates foundation models, knowledge graphs, and prompt engineering. FOKE aims to provide personalized, interactive, and explainable educational services by leveraging the strengths of these advanced techniques. In the following sections, we will introduce the key components and methodologies of FOKE in detail.

\subsection{Motivation}
Prompt engineering, an emerging paradigm in LLM application, has shown promising potential to address the aforementioned challenges. Prompt engineering refers to the process of designing and optimizing prompt templates, which are natural language instructions that guide LLMs to perform specific tasks \cite{WhiteJ}. By carefully crafting prompts, researchers have successfully adapted LLMs to a variety of downstream applications, such as visual question answering \cite{ShaoZ}, text classification \cite{MayerC}, and code augmentation \cite{AhmedT}. The success of prompt engineering can be attributed to its ability to leverage the knowledge and skills embedded in LLMs while providing a flexible and controllable interface for task customization.

In the context of intelligent education, prompt engineering holds great promise for enabling personalized, interactive, and explainable LLM-based services. By incorporating learner-specific information and domain knowledge into prompts, LLMs can generate tailored learning content and provide adaptive feedback to individual students. By designing conversation-oriented prompts, LLMs can engage in meaningful dialogue with learners and offer timely support and guidance. By injecting rationales and explanations into prompts, LLMs can produce interpretable and trustworthy educational suggestions and assessments. Therefore, the integration of prompt engineering and LLMs opens up new opportunities for revolutionizing intelligent education practices.

\subsection{Research Objectives}
Motivated by the potential of prompt engineering and LLMs in intelligent education, we propose FOKE, a novel framework that aims to achieve the following objectives:

\begin{itemize}
\item \textbf{Objective 1: Develop knowledge representation and user modeling techniques for personalized learning.} This objective includes (1) designing a hierarchical knowledge representation scheme to encode domain knowledge and support personalized learning paths; and (2) developing a multi-dimensional user profiling mechanism to capture learner characteristics, preferences, and styles, enabling adaptive educational services.

\item \textbf{Objective 2: Explore prompt engineering approaches for generating precise and interpretable learning guidance.} Under this objective, we aim to (1) design a structured prompt representation approach to elicit accurate and coherent feedback from LLMs; and (2) investigate the integration of knowledge representation and user modeling to facilitate personalized recommendation and adaptive instruction.

\item \textbf{Objective 3: Showcase the application scenarios and potential impact of FOKE in educational contexts.} This objective involves (1) demonstrating the effectiveness and advantages of FOKE through case studies in programming education, homework assessment, learning path planning, etc.; and (2) discussing the potential of FOKE in enhancing intelligent education practices and improving learning outcomes.
\end{itemize}
By achieving these objectives, FOKE is expected to provide a powerful and flexible framework for designing and deploying LLM-based intelligent education applications, while addressing the key challenges of personalization, interactivity, and interpretability.

\subsection{Contributions}
The main contributions of this work are summarized as follows:

\begin{enumerate}
    \item We propose FOKE, a novel framework that synergizes prompt engineering and LLMs to enable personalized, interactive, and explainable intelligent education services. Central to FOKE is a hierarchical knowledge forest, which represents domain knowledge as a tuple $\mathcal{KF} = (\mathcal{C}, \mathcal{R})$, where $\mathcal{C}$ denotes the set of concept nodes and $\mathcal{R}$ represents the set of relation edges.
    
    \item We design a multi-dimensional user profiling mechanism that integrates structured attributes, unstructured behaviors, and temporal trajectories to comprehensively model learner characteristics, which can be formulated as $\mathcal{UP} = (\mathcal{A}, \mathcal{B}, \mathcal{T})$, where $\mathcal{A}$, $\mathcal{B}$, and $\mathcal{T}$ correspond to the attribute, behavior, and trajectory dimensions, respectively. Based on the user profile, we propose a structured prompt representation scheme that incorporates goal specification, explanatory information, and feedback indicators to generate precise and informative learning guidance, with the prompt template represented as $\mathcal{PT} = (\mathcal{G}, \mathcal{E}, \mathcal{F})$, where $\mathcal{G}$, $\mathcal{E}$, and $\mathcal{F}$ denote the goal, explanation, and feedback components, respectively.
    
    \item We develop a graph embedding-based approach to fuse knowledge representation and user modeling, which learns low-dimensional vector representations for concepts and learners. The embedding learning process can be formulated as an optimization problem:
    \begin{equation}
        \min_{\mathbf{h}} \sum_{(c_i, c_j, r) \in \mathcal{D}} \mathcal{L}(\mathbf{h}_{c_i}, \mathbf{h}_{c_j}, \mathbf{h}_r),
    \end{equation}
    where $\mathbf{h}_{c_i}$, $\mathbf{h}_{c_j}$, and $\mathbf{h}_r$ are the embeddings of concepts $c_i$, $c_j$, and relation $r$, respectively, and $\mathcal{D}$ is the set of concept-relation triples. We designed three educational tasks to showcase the application of the FOKE framework: interactive programming education, personalized learning path planning, and intelligent homework assessment. Although a comprehensive evaluation of the framework is yet to be completed, AI educational products based on FOKE have already begun experiments and optimization in actual classroom teaching. By collecting feedback from students and teachers, we are continuously improving the system's performance and user experience. However, it is worth noting that currently in the field of intelligent education, there is a lack of standardized evaluation methods to compare different products, frameworks, or services, mainly due to the diversity of intelligent education systems, domain specificity, and differences in user needs. Establishing a comprehensive, fair, and effective set of evaluation criteria and methods is crucial for advancing the development of the intelligent education field. In our future work, we plan to collaborate with domain experts to provide more experience and suggestions for the evaluation of intelligent education systems while continuing to focus on improving and applying the FOKE framework, striving to achieve greater breakthroughs in personalization, interactivity, and interpretability, and contribute more innovative solutions to the field of intelligent education.
\end{enumerate}

\subsection{Organization}
The remainder of this paper is organized as follows. 
\textbf{Section \ref{sec:related}} reviews the related work on intelligent tutoring systems, knowledge representation, and personalized learning. 
\textbf{Section \ref{sec:method}} presents the proposed FOKE framework in detail, including the integration of foundation models and knowledge graphs, the construction of knowledge forests, the profiling of users, the engineering of interactive prompts, and the generation of explainable recommendations. 
\textbf{Section \ref{sec:application}} describes three representative application scenarios of FOKE, i.e., programming education, writing assessment, and learning path planning, and showcases the workflow and examples in each scenario. 
\textbf{Section \ref{sec:Scholar}} introduces the Scholar Hero system, an instantiation of FOKE in real-world educational settings, and reports its deployment, evaluation, and impact. 
\textbf{Section \ref{sec:conclusion}} concludes the paper and discusses future research directions.

\section{Related Work}
\label{sec:related}

In this section, we review the related work on AI in education, large language models, and prompt engineering, which form the theoretical and technical foundations of our proposed FOKE framework.

\subsection{AI in Education}
The application of AI in education has a long history, dating back to the 1970s when intelligent tutoring systems (ITS) were first introduced \cite{carbonell1970ai}. Since then, AI techniques have been extensively explored to enhance various aspects of education, such as personalized learning \cite{BernackiM}, adaptive assessment \cite{VieJ}, and learning analytics \cite{BanihashemS}. For example, Bayesian knowledge tracing \cite{YudelsonM} was proposed to model student knowledge acquisition and predict their performance. Reinforcement learning \cite{SinglaA} was employed to optimize instructional strategies and improve learning outcomes. Deep learning \cite{WarburtonK} was applied to capture complex learning patterns and enable intelligent tutoring interventions.

Despite the significant progress, existing AI-enhanced educational systems still face limitations in terms of adaptability, interactivity, and interpretability. Most systems rely on pre-defined rules or models, which may not effectively capture the diversity and dynamics of real-world educational scenarios. Moreover, the interaction between AI systems and learners is often restricted to simplistic forms, such as multiple-choice questions or short-answer responses, lacking the flexibility and naturalness of human-like communication. Furthermore, many AI models are black boxes, providing little explanatory information to help learners understand and trust the generated educational feedback. These limitations highlight the need for more advanced AI techniques that can enable personalized, interactive, and explainable education.

\subsection{Large Language Models}
The emergence of large language models (LLMs) has revolutionized the field of natural language processing (NLP) and opened up new opportunities for AI applications in various domains. LLMs, such as GPT-4 \cite{GPT4}, BERT \cite{BERT}, are deep neural networks pre-trained on massive text corpora, which can learn rich linguistic knowledge and perform a wide range of language tasks. For example, GPT-4 has shown impressive capabilities in language generation, translation, summarization, and question answering, surpassing human performance in some cases \cite{GPT4}. BERT has achieved state-of-the-art results on various NLP benchmarks, such as GLUE \cite{GLUE} and SQuAD \cite{SQuAD}, demonstrating its effectiveness in language understanding and reasoning.

The success of LLMs can be attributed to their ability to capture the intricate patterns and semantics of natural language from large-scale unsupervised learning. By training on diverse text data, LLMs can acquire a broad range of knowledge and skills, which can be transferred to downstream tasks with minimal fine-tuning. This transfer learning paradigm has greatly reduced the need for task-specific labeled data and enabled the development of generalizable and scalable AI systems. Moreover, the pre-training objectives of LLMs, such as masked language modeling \cite{SalazarJ} and permutation language modeling \cite{CuiY}, have been designed to encourage the models to learn contextual representations and long-range dependencies, which are essential for understanding and generating coherent and informative text.

However, the application of LLMs in education is still an emerging area that requires further exploration. Existing work has mainly focused on using LLMs for automated essay scoring \cite{RodriguezP}, question answering \cite{YasunagaM}. For example, Xue.J et al. \cite{XueJ} fine-tuned BERT on a large dataset of student essays and their corresponding scores, and achieved state-of-the-art performance on automated essay scoring. Abdelghani et al. \cite{AbdelghaniR} employed GPT-3 to generate diverse and informative answers to student questions, and demonstrated its potential for intelligent tutoring systems. While these studies have shown promising results, they have not fully exploited the potential of LLMs for interactive and explainable education, which is the focus of our FOKE framework.

\subsection{Prompt Engineering}
Prompt engineering is an emerging technique that aims to elicit desired behaviors from LLMs by designing and optimizing natural language prompts. Unlike traditional fine-tuning approaches that adapt LLMs to specific tasks through end-to-end training, prompt engineering keeps the LLM parameters fixed and instead crafts input prompts that can guide the model to perform the target task. This paradigm has several advantages, such as reducing the need for task-specific training data, enabling zero-shot or few-shot learning, and preserving the general knowledge and capabilities of LLMs.

The key idea of prompt engineering is to frame the task objective as a natural language prompt that can be easily understood and followed by LLMs. For example, to perform sentiment classification, a prompt template can be designed as "Classify the sentiment of the following text: \textit{[Text]}. The sentiment is \textit{[Sentiment]}.", where \textit{[Text]} is the input text and \textit{[Sentiment]} is the target sentiment label. By filling in the template with specific text instances and feeding them to LLMs, the models can generate the corresponding sentiment labels based on their language understanding and reasoning abilities. 


In the context of education, prompt engineering has the potential to enable personalized, interactive, and explainable LLM-based services. By incorporating learner-specific information and domain knowledge into prompts, LLMs can generate adaptive learning content and provide customized feedback. By designing conversation-oriented prompts, LLMs can engage in meaningful dialogue with learners and offer timely support and guidance. By injecting explanations and justifications into prompts, LLMs can produce interpretable and trustworthy educational suggestions and assessments. However, existing work on prompt engineering in education is still limited, which motivates us to develop the FOKE framework to fill this gap.

\section{FOKE Framework}
\label{sec:method}

In this section, we present the proposed FOKE framework in detail. We first provide an overview of the framework architecture and then elaborate on its key components, including the knowledge forest, user profiling, prompt representation, and graph embedding modules.
\begin{figure}
    \centering
    \includegraphics[width=0.8\linewidth]{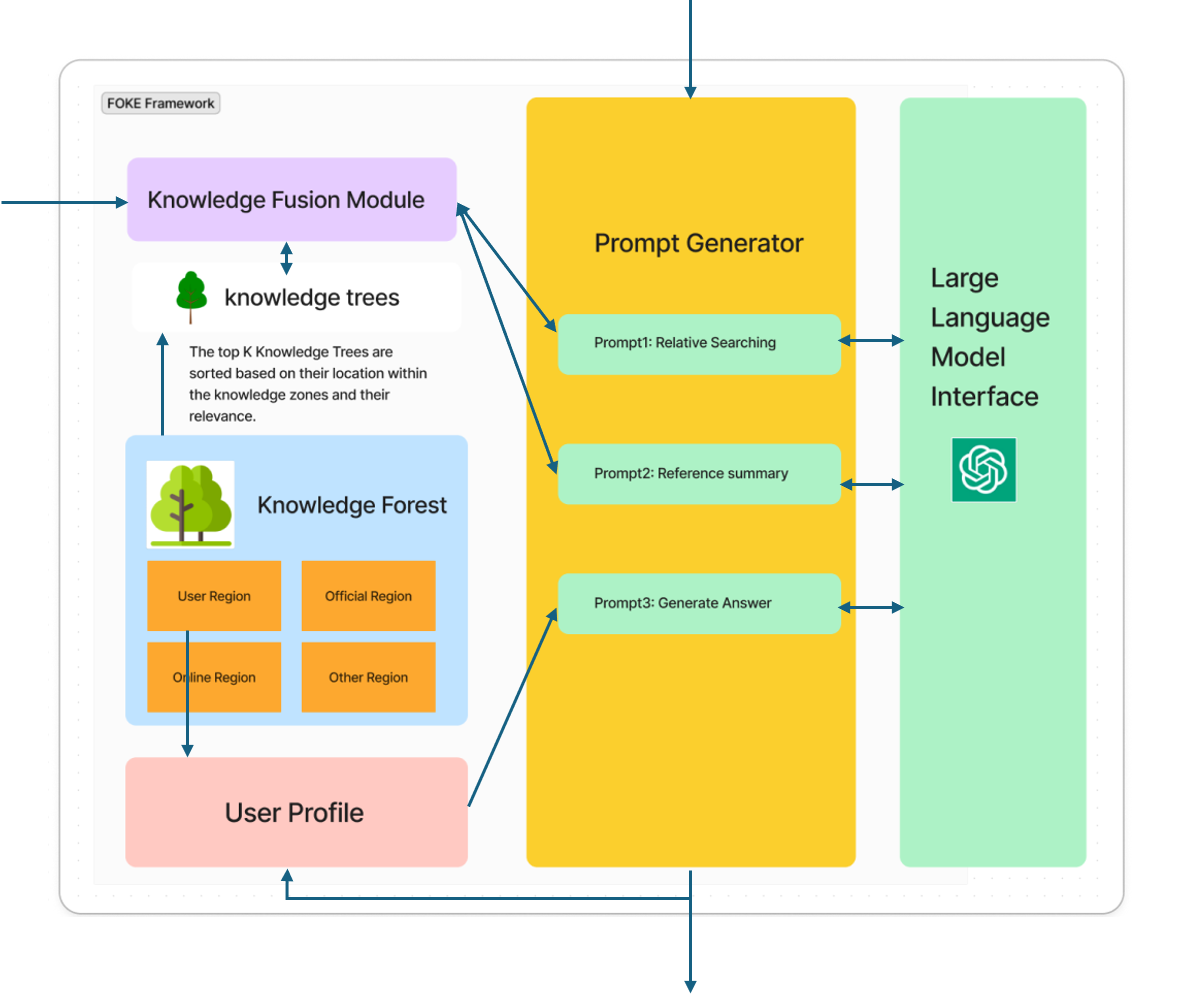}
    \caption{Overview of the FOKE Framework}
    \label{fig:enter-label}
\end{figure}
\subsection{Overview}
The framework takes as input a domain knowledge base, a user profile database, and a learning task specification, and generates personalized, interactive, and explainable learning guidance and feedback through the integration of knowledge representation, user modeling, prompt engineering, and graph embedding techniques.


Specifically, the domain knowledge base is first processed by the knowledge forest module to construct a hierarchical and structured representation of the key concepts and their relationships. The user profile database is then encoded by the user profiling module to capture the learners' attributes, behaviors, and temporal trajectories in a multi-dimensional manner. The learning task specification, along with the extracted knowledge and user representations, is fed into the prompt representation module to generate a set of structured prompts that guide the LLM to produce tailored and informative learning support. Finally, the graph embedding module fuses the knowledge and user representations through graph embedding techniques, which enables personalized recommendation and adaptive instruction based on the similarity and relevance of concepts and learners in the embedding space.

In the following subsections, we present each module of FOKE in more detail and formulate their key components and algorithms.

\subsection{Knowledge Forest}
The knowledge forest module introduces a novel and flexible approach to representing and organizing domain knowledge, addressing the limitations of traditional knowledge graphs. Unlike knowledge graphs, which typically focus on a single, static representation of knowledge, the knowledge forest allows for the dynamic integration of multiple knowledge trees, providing a more adaptable and efficient framework for knowledge acquisition and reasoning.

Formally, we define a knowledge forest as a collection of knowledge trees $\mathcal{KF} = \{\mathcal{T}_1, \mathcal{T}_2, \dots, \mathcal{T}_K\}$, where each tree $\mathcal{T}_k = (\mathcal{C}_k, \mathcal{R}_k, \mathcal{A}_k)$ represents a hierarchical structure of concepts $\mathcal{C}_k$, relations $\mathcal{R}_k$, and attributes $\mathcal{A}_k$. The key idea behind the knowledge forest is to enable the dynamic insertion and removal of knowledge trees based on the specific learning context and requirements, allowing for a more flexible and efficient knowledge representation compared to traditional knowledge graphs \cite{JiS}.

The construction of the knowledge forest follows a two-stage approach, similar to the single-tree case. In the first stage, we extract concepts and their attributes from the domain knowledge base using natural language processing techniques, such as named entity recognition \cite{LiJ}, relation extraction \cite{PawarS}, and semantic role labeling \cite{MàrquezL}. However, instead of constructing a single global concept set, we allow for the creation of multiple concept sets $\{\mathcal{C}_k\}_{k=1}^K$, each corresponding to a specific knowledge tree. This enables the knowledge forest to capture different aspects or subdomains of the knowledge base, providing a more fine-grained and context-specific representation.

In the second stage, we infer the semantic relationships within each knowledge tree $\mathcal{T}_k$ using graph mining and learning algorithms. However, the key difference lies in the time complexity of the inference process. In traditional knowledge graphs, the relation inference typically involves a global search over the entire concept space, resulting in a high time complexity of $O(|\mathcal{C}|^2)$, where $|\mathcal{C}|$ is the total number of concepts. In contrast, the knowledge forest allows for a local search within each knowledge tree, reducing the time complexity to $O(\sum_{k=1}^K |\mathcal{C}_k|^2)$, where $|\mathcal{C}_k|$ is the number of concepts in the $k$-th tree \cite{JiangC}. Since $|\mathcal{C}_k|$ is typically much smaller than $|\mathcal{C}|$, the knowledge forest enables a more efficient relation inference process.

Moreover, the knowledge forest introduces a novel tree-level inference mechanism to capture the relationships between different knowledge trees. Let $\mathbf{h}_k^{(L)}$ be the root embedding of the $k$-th knowledge tree, obtained by aggregating the embeddings of its concepts using a pooling function, such as mean or max pooling. We can then infer the tree-level relationships by measuring the similarity between the root embeddings:

\begin{equation}
    r(T_i, T_j) = \begin{cases}
        1, & \text{if } \mathrm{sim}(\mathbf{h}_i^{(L)}, \mathbf{h}_j^{(L)}) \geq \tau, \\
        0, & \text{otherwise},
    \end{cases}
\end{equation}

where $\mathrm{sim}(\cdot)$ is a similarity function, and $\tau$ is a predefined threshold. This tree-level inference allows the knowledge forest to capture the high-level relationships between different subdomains or aspects of knowledge, providing a more comprehensive and structured representation of the domain.

The dynamic nature of the knowledge forest enables efficient updates and adaptations to new knowledge. When a new knowledge tree $\mathcal{T}_{K+1}$ is discovered or constructed, it can be seamlessly inserted into the knowledge forest without the need to reconstruct the entire graph. Similarly, when a knowledge tree becomes obsolete or irrelevant, it can be easily removed from the forest. This dynamic property allows the knowledge forest to evolve and grow with the expanding knowledge base, making it a more scalable and maintainable solution compared to traditional knowledge graphs.

The knowledge forest also enables more efficient and targeted knowledge retrieval and reasoning. Given a query concept $c_q$, we can first identify the most relevant knowledge tree $\mathcal{T}_{k^*}$ based on the tree-level similarity:

\begin{equation}
    k^* = \underset{k \in \{1, 2, \dots, K\}}{\operatorname{argmax}} \, \mathrm{sim}(\mathbf{h}_q, \mathbf{h}_k^{(L)}),
\end{equation}

where $\mathbf{h}_q$ is the embedding of the query concept, and $\mathbf{h}_k^{(L)}$ is the root embedding of the $k$-th knowledge tree. Once the relevant tree is identified, we can perform a local search within $\mathcal{T}_{k^*}$ to retrieve the most relevant concepts and relations, avoiding the need to search the entire knowledge graph \cite{JiS}. This targeted retrieval process significantly reduces the search space and improves the efficiency of knowledge reasoning.

Furthermore, the knowledge forest enables personalized learning and recommendation by leveraging the hierarchical structure and tree-level relationships. By maintaining a student's knowledge state $\mathbf{s} = (s_1, s_2, \dots, s_K)$, where $s_k \in [0, 1]$ represents the student's mastery level of the $k$-th knowledge tree, we can recommend the most suitable knowledge tree for the student to explore next:

\begin{equation}
    k_{\text{next}} = \underset{k \in \{1, 2, \dots, K\}}{\operatorname{argmax}} \left(\sum_{i=1}^K r(T_i, T_k) \cdot s_i\right) \cdot (1 - s_k)
\end{equation}

where the first term measures the relevance of the $k$-th knowledge tree to the student's current knowledge state, and the second term ensures that the recommended tree is not yet fully explored by the student. This personalized recommendation strategy helps students navigate the knowledge forest efficiently and effectively, enhancing their learning experience and outcomes.

The knowledge forest also facilitates the integration of multi-modal learning resources, such as text, images, and videos, by associating them with the relevant concepts and relations in the knowledge trees \cite{YagerR}. This multi-modal representation enhances the richness and diversity of the learning content, catering to different learning styles and preferences.

Moreover, the knowledge forest supports explainable reasoning and decision-making by providing a transparent and interpretable representation of the knowledge structure. The hierarchical organization of concepts and the explicit modeling of semantic relationships enable the generation of human-understandable explanations for the recommended learning paths and resources.

In summary, the knowledge forest module introduces a dynamic, flexible, and efficient approach to representing and organizing domain knowledge. By allowing for the integration of multiple knowledge trees and enabling tree-level inference, personalized recommendation, multi-modal learning, and explainable reasoning, the knowledge forest addresses the limitations of traditional knowledge graphs and provides a more scalable and adaptable solution for knowledge acquisition and reasoning in educational applications.

\subsection{User Profiling}
The user profiling module aims to capture the learners' characteristics and preferences in a multi-dimensional manner to enable personalized and adaptive learning support. We define a user profile as a tuple $\mathcal{UP} = (\mathcal{A}, \mathcal{B}, \mathcal{T})$, where $\mathcal{A}$, $\mathcal{B}$, and $\mathcal{T}$ denote the attribute, behavior, and trajectory dimensions, respectively.

The attribute dimension $\mathcal{A}$ represents the learner's static and demographic information, such as age, gender, education level, and learning style. The behavior dimension $\mathcal{B}$ captures the learner's dynamic and interactive patterns, such as learning activities, quiz performance, and forum participation. The trajectory dimension $\mathcal{T}$ models the learner's temporal and sequential characteristics, such as learning paths, knowledge states, and skill acquisition.

To construct the user profiles, we propose a multi-view representation learning approach. For each dimension, we extract the relevant features from the user profile database and learn a low-dimensional vector representation using techniques such as matrix factorization \cite{MnihA}, deep learning \cite{WideDeep}, and sequential modeling \cite{KangW}. We then fuse the learned representations from different dimensions using attention mechanisms \cite{LuJ} or graph convolutional networks \cite{YuJ} to obtain a unified user embedding.

Formally, let $\mathbf{a}_u$, $\mathbf{b}_u$, and $\mathbf{t}_u$ denote the attribute, behavior, and trajectory embeddings of user $u$, respectively. The unified user embedding $\mathbf{h}_u$ can be obtained as:
\begin{equation}
    \mathbf{h}_u = f(\mathbf{a}_u, \mathbf{b}_u, \mathbf{t}_u; \Theta),
\end{equation}
where $f$ is the fusion function with parameters $\Theta$. For example, an attention-based fusion function can be defined as:
\begin{align}
    \alpha_a, \alpha_b, \alpha_t &= \text{softmax}(\mathbf{w}_a^\top\mathbf{a}_u, \mathbf{w}_b^\top\mathbf{b}_u, \mathbf{w}_t^\top\mathbf{t}_u),\\
    \mathbf{h}_u &= \alpha_a\mathbf{a}_u + \alpha_b\mathbf{b}_u + \alpha_t\mathbf{t}_u,
\end{align}
where $\mathbf{w}_a$, $\mathbf{w}_b$, and $\mathbf{w}_t$ are the attention weight vectors for the attribute, behavior, and trajectory dimensions, respectively.

\subsection{Prompt Representation}
The prompt representation module aims to generate structured and informative prompts that can guide the LLM to produce personalized and explainable learning support. We define a prompt as a tuple $\mathcal{PT} = (\mathcal{G}, \mathcal{E}, \mathcal{F})$, where $\mathcal{G}$, $\mathcal{E}$, and $\mathcal{F}$ denote the goal, explanation, and feedback components, respectively.

The goal component $\mathcal{G}$ specifies the learning objective and task requirements, such as knowledge acquisition, problem-solving, or skill practice. The explanation component $\mathcal{E}$ provides the relevant background information and reasoning process to help learners understand the key concepts and their relationships. The feedback component $\mathcal{F}$ offers personalized suggestions and assessments based on the learner's performance and characteristics.

To generate the structured prompts, we propose a template-based approach that leverages the knowledge forest and user profile representations. Specifically, for each learning task, we first retrieve the relevant concepts and relations from the knowledge forest based on the task specification. We then instantiate the prompt template with the retrieved knowledge and the learner's attributes and preferences. Finally, we optimize the prompt template and instantiation using techniques such as reinforcement learning \cite{DengM} to improve the quality and diversity of the generated prompts.

Formally, let $\mathcal{KF}_t \subseteq \mathcal{KF}$ denote the subset of the knowledge forest relevant to task $t$, and $\mathbf{h}_u$ denote the user embedding of learner $u$. The prompt generation process can be formulated as:
\begin{align}
\mathcal{G}_t &= g(\mathcal{KF}_t, \mathbf{h}_u; \Theta_g), \\
\mathcal{E}_t &= e(\mathcal{KF}_t, \mathbf{h}_u; \Theta_e), \\
\mathcal{F}_t &= f(\mathcal{KF}_t, \mathbf{h}_u; \Theta_f), \\
\mathcal{PT}_t &= (\mathcal{G}_t, \mathcal{E}_t, \mathcal{F}_t),
\end{align}
where $g$, $e$, and $f$ are the generation functions for the goal, explanation, and feedback components, respectively, with parameters $\Theta_g$, $\Theta_e$, and $\Theta_f$.

For example, a template for the goal component can be defined as:
\begin{quote}
Acquire knowledge about \texttt{[Concept]} and its relations to \texttt{[RelatedConcepts]} to solve problems of type \texttt{[ProblemType]}.
\end{quote}
where \texttt{[Concept]}, \texttt{[RelatedConcepts]}, and \texttt{[ProblemType]} are slots to be filled with the retrieved knowledge and user information.

The template can be instantiated as:
\begin{quote}
Acquire knowledge about \textit{dynamic programming} and its relations to \textit{recursion} and \textit{memoization} to solve problems of type \textit{optimization}, given your interest in \textit{algorithms} and \textit{competitive programming}.
\end{quote}

The quality of the generated prompt $\mathcal{PT}t$ can be evaluated based on metrics such as relevance, coherence, and diversity, and optimized using reinforcement learning objectives such as:
\begin{equation}
\max{\Theta_g, \Theta_e, \Theta_f} \mathbb{E}_{\mathcal{PT}_t \sim p(\cdot|\mathcal{KF}_t, \mathbf{h}_u)}[R(\mathcal{PT}_t)],
\end{equation}
where $R(\mathcal{PT}_t)$ is the reward function that measures the quality and effectiveness of the prompt.

\subsection{Graph Embedding}
The graph embedding module aims to learn a unified representation of the knowledge forest and user profiles to enable personalized recommendation and adaptive instruction. We propose a graph embedding approach that learns low-dimensional vector representations of concepts and users based on their structural and semantic relationships.

Formally, let $\mathcal{G} = (\mathcal{V}, \mathcal{E})$ denote the graph constructed from the knowledge forest $\mathcal{KF}$ and user profiles ${\mathcal{UP}_u}$, where $\mathcal{V} = \mathcal{C} \cup \mathcal{U}$ is the set of nodes (concepts and users) and $\mathcal{E}$ is the set of edges (relations and interactions). The goal is to learn a function $f: \mathcal{V} \rightarrow \mathbb{R}^d$ that maps each node to a $d$-dimensional vector, such that the graph structure and semantics are preserved in the embedding space.

We adopt a graph neural network (GNN) approach to learn the node embeddings. At each layer $l$, the embedding of node $v$ is updated as:
\begin{align}
\mathbf{m}_v^{(l)} &= \text{AGG}({\mathbf{h}_u^{(l-1)} : u \in \mathcal{N}(v)}),\
\mathbf{h}_v^{(l)} &= \text{UPDATE}(\mathbf{h}_v^{(l-1)}, \mathbf{m}_v^{(l)}),
\end{align}
where $\mathbf{h}_v^{(l)}$ is the embedding of node $v$ at layer $l$, $\mathcal{N}(v)$ is the set of neighbors of $v$, and $\text{AGG}$ and $\text{UPDATE}$ are the aggregation and update functions, respectively.

The GNN is trained using a combination of supervised and unsupervised objectives. For supervised learning, we use the node embeddings to predict the node labels (e.g., concept categories or user clusters) based on a cross-entropy loss:
\begin{equation}
\mathcal{L}s = -\sum{v \in \mathcal{V}L} \sum{c=1}^C y_{vc} \log \hat{y}{vc},
\end{equation}
where $\mathcal{V}L \subseteq \mathcal{V}$ is the set of labeled nodes, $C$ is the number of classes, $y{vc}$ is the true label of node $v$ for class $c$, and $\hat{y}{vc}$ is the predicted probability.

For unsupervised learning, we use a contrastive loss \cite{WangF} to enforce the similarity of embeddings for neighboring nodes and the dissimilarity of embeddings for random pairs of nodes:
\begin{equation}
\mathcal{L}u = -\sum{(u,v) \in \mathcal{E}} \log \frac{\exp(\text{sim}(\mathbf{h}_u, \mathbf{h}v))}{\sum{v' \in \mathcal{V}} \exp(\text{sim}(\mathbf{h}u, \mathbf{h}{v'}))},
\end{equation}
where $\text{sim}(\cdot,\cdot)$ is a similarity function such as cosine similarity or dot product.

The overall training objective is a weighted combination of the supervised and unsupervised losses:
\begin{equation}
\mathcal{L} = \lambda_s \mathcal{L}_s + \lambda_u \mathcal{L}_u,
\end{equation}
where $\lambda_s$ and $\lambda_u$ are the weighting coefficients.

After training, the learned node embeddings can be used for various downstream tasks, such as concept recommendation, prerequisite prediction, and learning path generation, based on the similarity and relevance of the embeddings.

\section{Application Scenarios}
\label{sec:application}

To demonstrate the practical value and wide applicability of the proposed FOKE framework, we showcase its potential in three representative educational scenarios: interactive programming education, personalized learning path planning, and intelligent writing assessment. In this section, we describe the workflow and key functionalities of FOKE in each scenario, and provide illustrative examples to show its effectiveness in generating personalized and explainable learning support.

\begin{figure}
    \centering
    \includegraphics[width=0.8\linewidth]{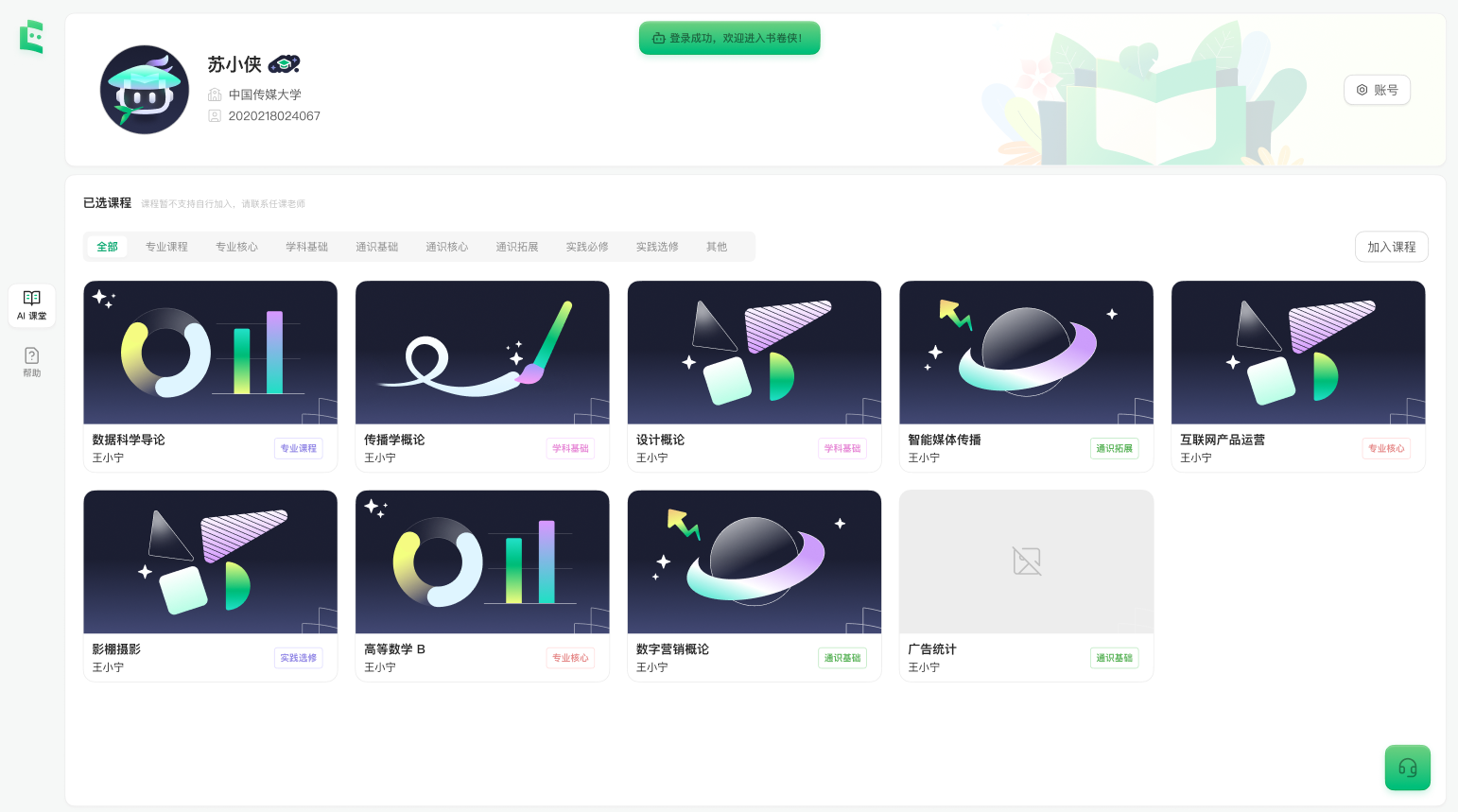}
    \caption{Scholar Hero: The Application of the FOKE framework.}
    \label{fig:sha}
\end{figure}

\subsection{Interactive Programming Education}
Programming education is a challenging yet important domain that requires learners to acquire complex concepts and skills. FOKE can be applied to provide interactive and adaptive programming tutorials and feedback to enhance the learning experience and outcome. 

The workflow of FOKE in programming education is as follows:
\begin{enumerate}
    \item The knowledge forest of programming concepts and skills is constructed from various learning resources such as textbooks, online courses, and coding platforms.
    \item The user profile of each learner is built based on their background, performance, and learning activities in programming.
    \item For each programming task or project, FOKE generates a set of personalized prompts that guide the learner through the problem-solving process, such as understanding the requirements, designing the solution, and debugging the code.
    \item The learner interacts with the prompts and receives adaptive feedback and suggestions based on their input and progress.
    \item The knowledge forest and user profile are updated based on the learner's performance and feedback to enable continuous optimization and personalization.
\end{enumerate}

For example, suppose a learner is working on a Python project to implement a recommendation system. FOKE may generate the following prompts to guide the learner:
\begin{quote}
    \textbf{Prompt 1:} Let's break down the project requirements into smaller tasks. Based on your experience with Python and machine learning, what are the key steps you would take to build a recommendation system? \\
    \textbf{Prompt 2:} Great insights! To implement the collaborative filtering algorithm, you need to first construct the user-item interaction matrix. Here are a few examples of how to represent the matrix using NumPy or Pandas. Which representation do you think is more suitable for your project and why? \\
    \textbf{Prompt 3:} Your code looks good so far. I noticed that you used a nested loop to calculate the user similarity scores, which may be inefficient for large datasets. Have you considered using matrix operations or libraries like SciPy to speed up the computation? Here are a few suggestions...
\end{quote}

Through the interactive and explainable prompts, FOKE can help learners deepen their understanding of programming concepts, improve their problem-solving skills, and receive timely and actionable feedback. The personalization and adaptivity of the prompts also cater to the diverse needs and preferences of learners, making the learning process more engaging and effective.

\subsection{Personalized Learning Path Planning}
Learning path planning is a critical task in education that aims to recommend a sequence of courses or learning activities for learners to achieve their learning goals. FOKE can be applied to generate personalized learning paths based on the learner's knowledge state, learning style, and career aspiration.

The application mechanism of FOKE in learning path planning is as follows:
\begin{enumerate}
    \item The knowledge forest of course prerequisites, topics, and learning outcomes is constructed from the course catalog and syllabus.
    \item The user profile of each learner is built based on their academic record, learning preferences, and career goals.
    \item FOKE generates a set of candidate learning paths based on the knowledge forest and the learner's profile, using techniques such as prerequisite-based filtering, collaborative filtering, and goal-based reasoning.
    \item The candidate learning paths are ranked and recommended to the learner based on their predicted performance, satisfaction, and alignment with their goals.
    \item The learner explores and selects the learning paths, and provides feedback and preference to update their profile and refine the recommendation.
\end{enumerate}

For instance, consider a college student majoring in computer science who wants to become a data scientist. FOKE may recommend the following learning path:
\begin{quote}
    \textbf{Step 1:} Take the "Introduction to Data Science" course to learn the basics of data manipulation, visualization, and machine learning. This course aligns well with your background in programming and statistics, and will provide a solid foundation for more advanced topics. \\
    \textbf{Step 2:} Enroll in the "Database Systems" and "Data Mining" courses to deepen your knowledge of data management and analysis. These courses build upon the concepts you learned in the introductory course and are highly relevant to your career goal. \\
    \textbf{Step 3:} Gain practical experience by participating in data science projects or internships. Based on your performance in the previous courses and your interest in healthcare, we recommend joining the "Medical Image Analysis" project led by Prof. Smith. This project will help you apply your skills to real-world problems and enhance your resume.
\end{quote}

By providing personalized and explainable learning path recommendations, FOKE can help learners navigate the complex course space, discover their interests and strengths, and achieve their learning and career goals more effectively and efficiently. The interactive and dynamic nature of the recommendation process also allows learners to actively shape their learning experience and adapt to their changing needs and preferences.

\subsection{Intelligent Homework Assessment}
Writing assessment(A Example of Homework Assessment) is a time-consuming and subjective task that requires teachers to provide detailed and constructive feedback to help students improve their writing skills. FOKE can be applied to automate the writing assessment process and generate objective, comprehensive, and actionable feedback.

The technical workflow of FOKE in writing assessment is as follows:
\begin{enumerate}
    \item The knowledge forest of writing aspects, criteria, and examples is constructed from the rubrics, writing prompts, and exemplar essays.
    \item The user profile of each student is built based on their writing history, performance, and feedback.
    \item For each writing assignment, FOKE generates a set of prompts to elicit the student's self-assessment and reflection on their writing strengths and weaknesses.
    \item The student's essay is analyzed using natural language processing techniques to identify the key aspects such as coherence, argument, and grammar. The results are compared with the writing criteria and examples in the knowledge forest to generate an evaluation report.
    \item The evaluation report is presented to the student in an interactive format, with detailed feedback, suggestions, and examples for each aspect of their writing. The student can explore the feedback, ask questions, and revise their essay accordingly.
\end{enumerate}

For example, suppose a student submits an argumentative essay on the topic of "social media and mental health". FOKE may generate the following feedback report:
\begin{quote}
    \textbf{Aspect 1: Thesis Statement}
    Your thesis statement is clear and concise, stating your main argument that social media has negative impacts on mental health. However, you could make it more specific by mentioning the key points you will use to support your argument, such as addiction, self-esteem, and cyberbullying.

    \textbf{Aspect 2: Evidence and Examples}
    You provided some relevant evidence to support your argument, such as the study showing the correlation between social media use and depression. However, your examples are mostly anecdotal and lack credibility. Consider using more research findings or expert opinions to strengthen your case.

    \textbf{Aspect 3: Counterargument and Rebuttal}
    You mentioned a counterargument that social media can also have positive effects on mental health, such as connecting people and providing support. However, your rebuttal is weak and does not effectively address the counterargument. You could argue that the negative effects outweigh the positive ones, or provide evidence to show the limitations of the counterargument.
\end{quote}

By generating objective, detailed, and actionable feedback on various aspects of writing, FOKE can help students identify their strengths and weaknesses, understand the criteria and expectations, and improve their writing skills through iterative revision and reflection. The interactive and personalized nature of the feedback also facilitates student-teacher communication and collaboration, making the writing assessment process more efficient, effective, and engaging.


\section{Scholar Hero: Empowering Education with FOKE}
\label{sec:Scholar}

Scholar Hero is a pioneering AI-driven educational application that embodies the core principles and innovations of the FOKE framework. By seamlessly integrating advanced AI agents and RAG (Retrieval-Augmented Generation) technologies, Scholar Hero aims to revolutionize the traditional education landscape and bring tangible benefits to every student's learning journey and campus life.

\subsection{System Architecture and Workflow}

The architecture of Scholar Hero is built upon the FOKE framework, leveraging its hierarchical knowledge representation, multi-dimensional user profiling, and structured prompt generation. 

The system consists of three main components: 1) Knowledge Base Construction, 2) User Profiling and Personalization, and 3) Interactive Learning Services.

In the Knowledge Base Construction phase, Scholar Hero employs advanced data mining and knowledge extraction techniques to build a comprehensive and hierarchical knowledge base covering a wide range of educational topics and campus-related information. The knowledge base is continuously updated and expanded to ensure its relevance and accuracy.

The User Profiling and Personalization component is responsible for creating and maintaining multi-dimensional user profiles based on students' learning preferences, academic performance, interests, and behavioral data. These profiles enable Scholar Hero to deliver highly personalized content, recommendations, and services tailored to each individual student's needs.

The Interactive Learning Services component is the core of Scholar Hero, powered by AI agents and RAG technology. It provides a wide array of intelligent educational services, including personalized course recommendations, interactive tutoring, intelligent homework assistance, and AI-generated learning materials. The AI agents engage in natural language interactions with students, answering their questions, providing explanations, and offering guidance and support throughout their learning process.

\subsection{Unique Features and Innovations}

Scholar Hero distinguishes itself from traditional educational applications through its unique features and innovations, which are deeply rooted in the FOKE framework:

\begin{itemize}
    \item \textbf{Personalized Learning Paths}: Scholar Hero generates dynamic and personalized learning paths for each student based on their knowledge level, learning style, and goals. The system continuously adapts and optimizes these paths as students progress, ensuring a truly personalized learning experience.

    \item \textbf{Interactive and Engaging Learning}: The AI agents in Scholar Hero provide interactive and engaging learning experiences through natural language conversations, real-time feedback, and multimedia content. Students can ask questions, seek clarifications, and receive instant support, fostering a more active and immersive learning environment. Figure \ref{fig:interactive_learning} demonstrates a student interacting with an AI agent in Scholar Hero.

    \begin{figure}
        \centering
        \includegraphics[width=0.8\linewidth]{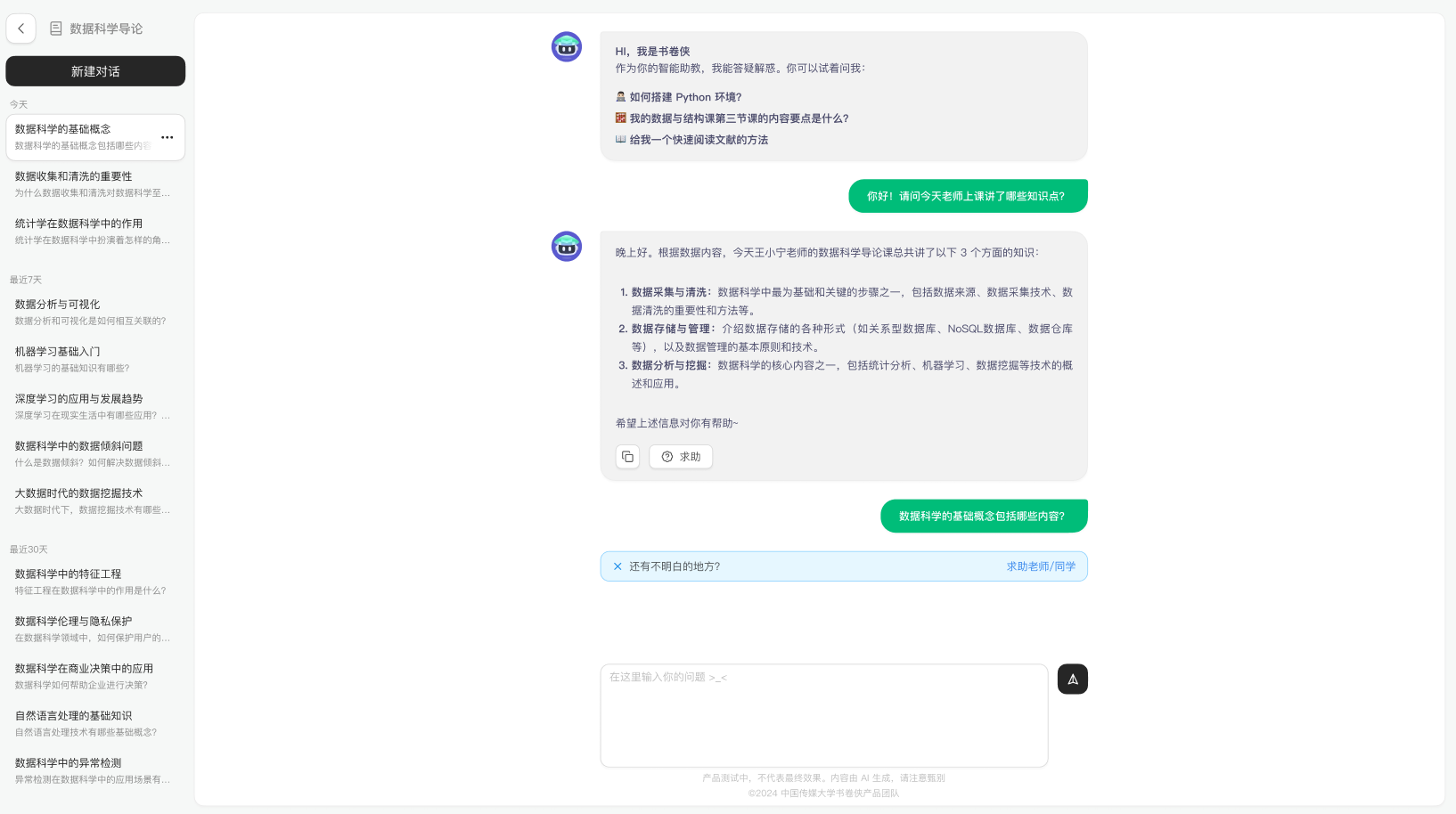}
        \caption{A student interacting with an AI agent in Scholar Hero}
        \label{fig:shAIl}
    \end{figure}
    \begin{figure}[htbp]
        \centering
        \includegraphics[width=0.8\linewidth]{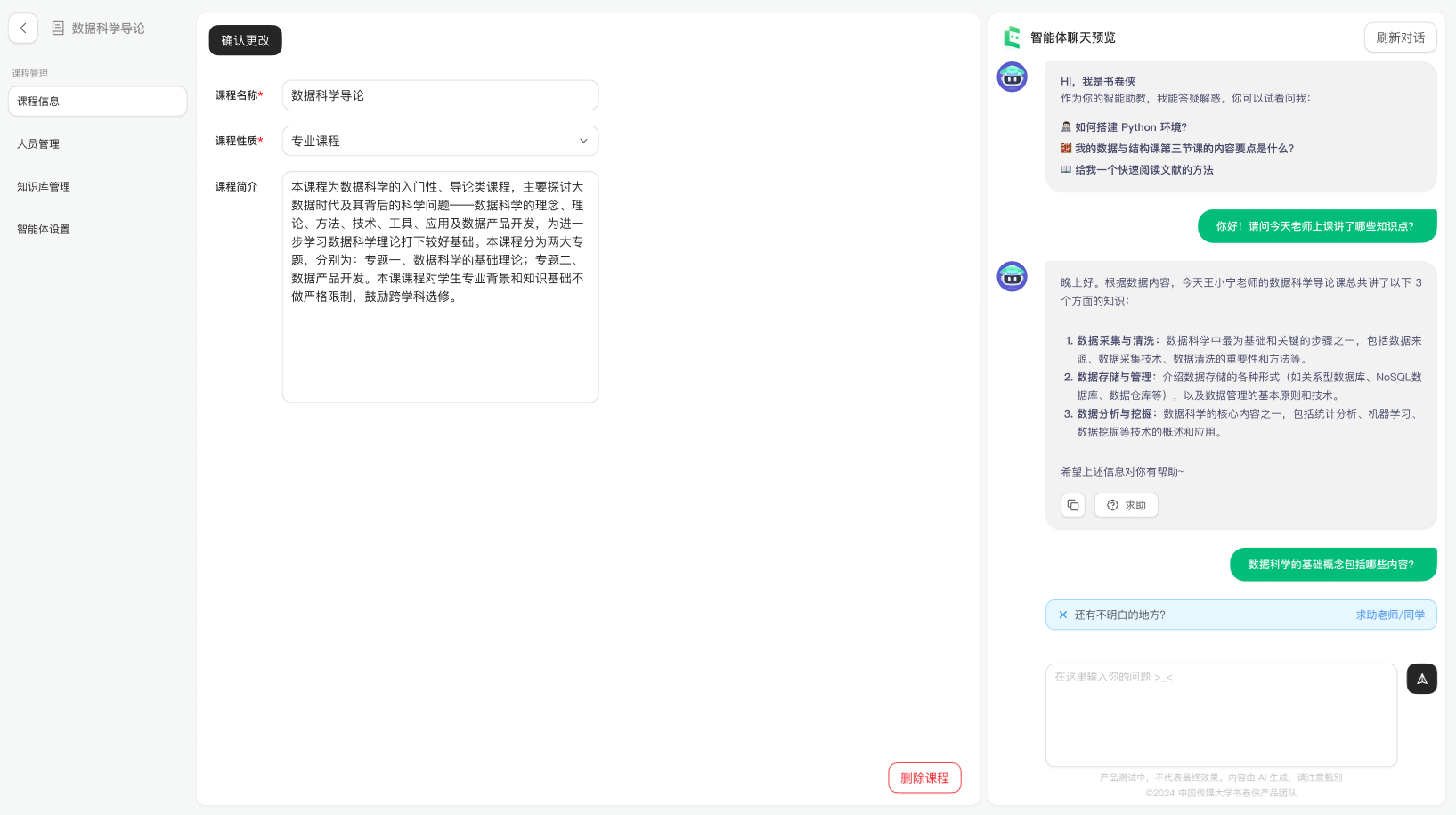}
        \caption{teacher setting an AI agent in Scholar Hero.}
        \label{fig:interactive_learning}
    \end{figure}
    \begin{figure}
        \centering
        \includegraphics[width=0.8\linewidth]{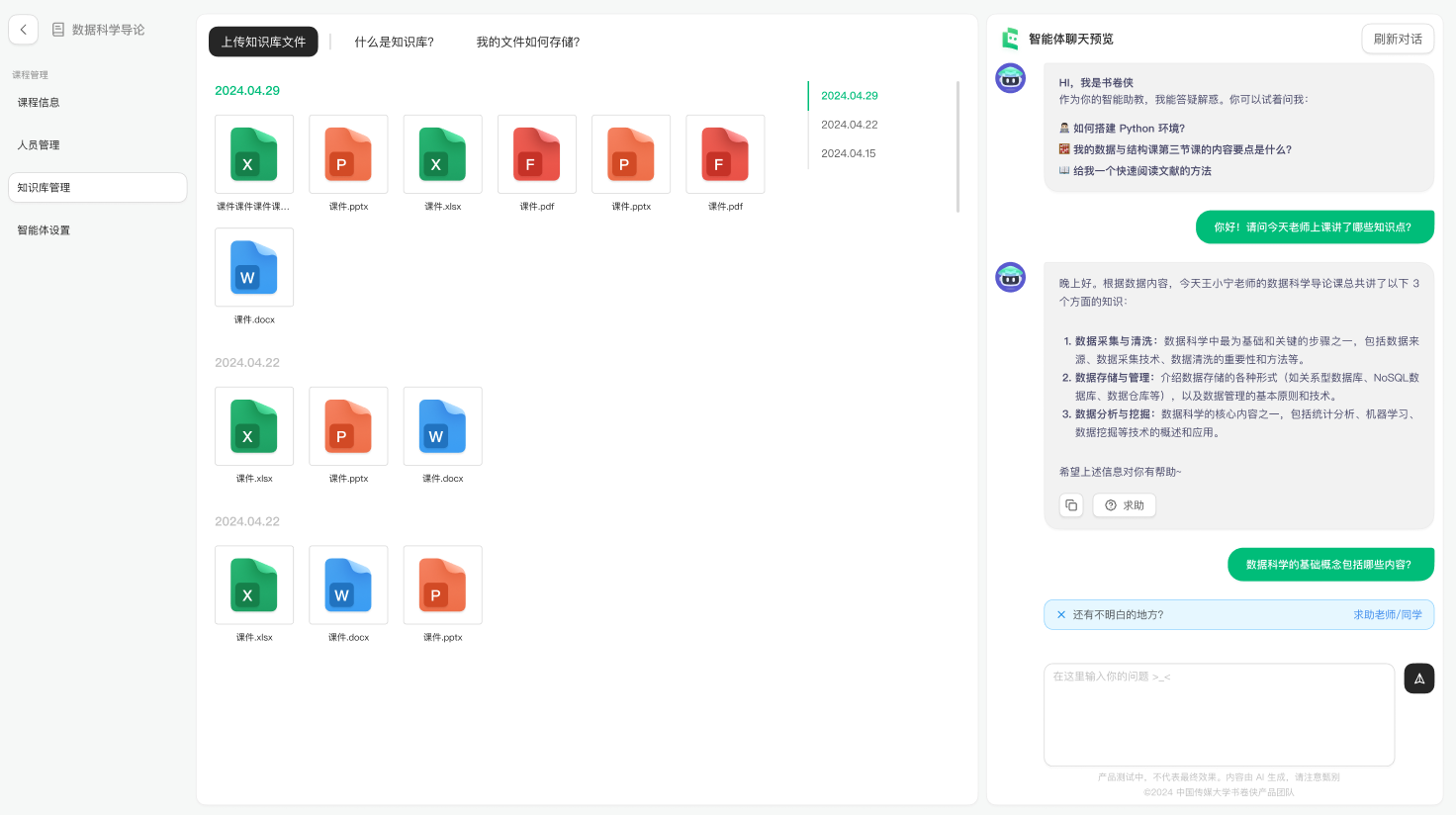}
        \caption{teacher setting the Knowledge Forest of AI agent.}
        \label{fig:shKB}
    \end{figure}

    \item \textbf{Intelligent Homework Assistance}: Scholar Hero offers intelligent homework assistance by leveraging the knowledge base and AI agents. It can automatically generate explanations, provide step-by-step guidance, and offer personalized feedback on students' assignments, helping them to identify their strengths and weaknesses and improve their understanding of the subject matter.
    
    \item \textbf{Seamless Integration with Campus Life}: Scholar Hero goes beyond academic learning by integrating various aspects of campus life into its services. From finding the best food options on campus to managing schedules and extracurricular activities, Scholar Hero serves as a comprehensive and intelligent companion for students throughout their university journey.
    
    \item \textbf{Continuous Learning and Improvement}: Scholar Hero is designed to continuously learn and improve through user interactions and feedback. The system employs advanced machine learning algorithms to refine its knowledge base, user profiles, and recommendation models, ensuring its services remain accurate, relevant, and up-to-date.
\end{itemize}

\subsection{Real-World Impact and Future Directions}

Scholar Hero has been successfully deployed in several pilot universities, receiving enthusiastic feedback from students and educators alike. The application has demonstrated significant potential in enhancing student engagement, improving academic performance, and promoting a more personalized and effective learning experience. 

As we continue to develop and expand Scholar Hero, we envision a future where AI-driven educational applications become an integral part of the educational ecosystem. By leveraging the power of the FOKE framework and the lessons learned from Scholar Hero, we aim to create more intelligent, adaptive, and inclusive educational solutions that cater to the diverse needs of learners worldwide.

Furthermore, we plan to conduct extensive user studies and evaluations to assess the long-term impact of Scholar Hero on student learning outcomes, satisfaction, and overall educational experience. The insights gained from these studies will not only inform the future development of Scholar Hero but also contribute to the broader research community in AI-driven education.

In conclusion, Scholar Hero stands as a testament to the transformative potential of the FOKE framework in reshaping the educational landscape. By harnessing the power of AI, we are paving the way for a new era of personalized, engaging, and effective learning experiences that empower students to thrive in their academic and personal lives.

\section{Conclusion and Future Work}
\label{sec:conclusion}

In this work, we proposed a novel framework named FOKE that integrates foundation models and knowledge graphs for intelligent and personalized education. FOKE leverages the power of large language models and domain knowledge to generate adaptive learning support, such as interactive prompts, personalized feedback, and explainable recommendations. Through a comprehensive analysis of three representative application scenarios, we demonstrated the potential of FOKE in enhancing the effectiveness, efficiency, and engagement of various educational tasks, such as programming education, learning path planning, and writing assessment.

The key contributions of this work include:
\begin{itemize}
    \item A flexible and extensible framework that incorporates foundation models and knowledge graphs to capture the complex relationships among learning content, learner characteristics, and pedagogical strategies;
    \item A set of innovative techniques, such as prompt engineering, knowledge forest construction, and explainable recommendation, to generate personalized and interactive learning support;
    \item A systematic analysis of the application scenarios, workflows, and examples of FOKE in three important educational domains, showcasing its wide applicability and practical value.
\end{itemize}

Our work explores a promising direction in developing next-generation AI-powered educational systems that can adapt to the diverse needs and preferences of learners, and provide intelligent and engaging learning experiences at scale. The generic and modular design of FOKE also makes it potentially applicable to other educational contexts, such as corporate training, online learning, and lifelong learning. By leveraging the rapid advancement of foundation models and knowledge graphs, FOKE can continuously evolve and improve its capabilities in understanding learners, knowledge, and pedagogy.

As a concrete application of the FOKE framework, we have developed an intelligent educational system named "Scholar Hero". Scholar Hero is designed to provide personalized and interactive learning support for various subjects and domains, with a special focus on data science education. Currently, Scholar Hero is being piloted in an introductory data science course at the Communication University of China, where it serves as a virtual teaching assistant to provide adaptive learning materials, formative assessments, and real-time feedback to students.

The development and deployment of Scholar Hero is an ongoing process, with continuous iterations and improvements based on the feedback and data collected from the pilot study. Preliminary results show that Scholar Hero has the potential to improve students' learning outcomes, engagement, and satisfaction, by providing them with tailored and timely support that caters to their individual needs and progress. We plan to conduct more rigorous and large-scale evaluations of Scholar Hero in the future, and explore its application in other educational settings and domains.

Despite the promising progress, our work still has several limitations and challenges. First, the knowledge forest and user profiles in Scholar Hero are currently constructed based on limited and domain-specific data, which may not capture the full complexity and diversity of learners and subjects. Second, the prompt engineering and recommendation techniques used in Scholar Hero are still rudimentary and need to be further optimized for efficiency and effectiveness. Third, the long-term impact and generalizability of Scholar Hero need to be validated through longitudinal and cross-domain studies.

Therefore, we plan to pursue the following directions in our future work:
\begin{itemize}
    \item Expand the knowledge forest and user profiling capabilities of Scholar Hero by incorporating more diverse and large-scale educational data, such as learning resources, assessments, and student records;
    \item Develop more advanced and scalable techniques for prompt engineering and recommendation, leveraging the state-of-the-art research in natural language processing, knowledge representation, and recommender systems;
    \item Conduct more comprehensive and long-term evaluations of Scholar Hero, both in the current pilot study and in other educational contexts, to assess its effectiveness, usability, and generalizability;
    \item Investigate the pedagogical and ethical implications of using AI-powered educational systems like Scholar Hero, such as the alignment with learning theories, the fairness and transparency of the generated support, and the impact on teacher-student relationships.
\end{itemize}

We believe that our work on the FOKE framework and its application in Scholar Hero represents an important step towards realizing the full potential of AI in education. By integrating foundation models, knowledge graphs, and educational domain knowledge, we can create intelligent and personalized learning experiences that can benefit learners of diverse backgrounds and abilities. We hope that our work can inspire more research and development efforts in this direction, and contribute to the advancement of AI-powered education.

%
%
%
\bibliographystyle{splncs04}
%

\end{document}